\documentclass[a4paper,12pt]{article}
\usepackage{jheppub,esint,shuffle,psfrag}
\usepackage[utf8]{inputenc}

\usepackage{epsfig,amssymb,amsmath,psfrag,subfigure,rotate,color,wasysym,xcolor}
\usepackage[bbgreekl]{mathbbol}

\allowdisplaybreaks
\newcommand{\insertfig}[2]{\includegraphics[width=#1cm]{#2}}

\DeclareSymbolFontAlphabet{\mathbbm}{bbold}
\DeclareSymbolFontAlphabet{\mathbb}{AMSb}%

\def\XXint#1#2#3{{\setbox0=\hbox{$#1{#2#3}{\int}$ }
\vcenter{\hbox{$#2#3$ }}\kern-.6\wd0}}

\def \be  {\begin{equation}}
\def \ee  {\end{equation}}
\def \ba  {\begin{eqnarray}}
\def \ea  {\end{eqnarray}}
\def \baa {\begin{eqnarray*}}
\def \eaa {\end{eqnarray*}}

\def \lab #1 {\label{#1}}

\def\d{\hbox{{d}\kern-.20em\hbox{l}}}

\def \matrix #1 {\left(\begin{array}{cc} #1 \end{array}\right)}

\newcommand{\bit}[1]{\mbox{\boldmath$#1$}}
\def\1{\hbox{{1}\kern-.25em\hbox{l}}}

\makeatletter

\input pdf-trans
\newbox\qbox
\def\usecolor#1{\csname\string\color@#1\endcsname\space}
\newcommand\bordercolor[1]{\colsplit{1}{#1}}
\newcommand\fillcolor[1]{\colsplit{0}{#1}}
\newcommand\outline[1]{\leavevmode%
  \def\maltext{#1}%
  \setbox\qbox=\hbox{\maltext}%
  \boxgs{Q q 2 Tr \thickness\space w \fillcol\space \bordercol\space}{}%
  \copy\qbox%
}
\makeatother
\newcommand\colsplit[2]{\colorlet{tmpcolor}{#2}\edef\tmp{\usecolor{tmpcolor}}%
  \def\tmpB{}\expandafter\colsplithelp\tmp\relax%
  \ifnum0=#1\relax\edef\fillcol{\tmpB}\else\edef\bordercol{\tmpC}\fi}
\def\colsplithelp#1#2 #3\relax{%
  \edef\tmpB{\tmpB#1#2 }%
  \ifnum `#1>`9\relax\def\tmpC{#3}\else\colsplithelp#3\relax\fi
}
\bordercolor{black}
\fillcolor{white}
\def\thickness{.3}

\def\1{\mathbbm{1}}

\title{Off-shell form factor in $\mathcal{N}$=4 sYM at three loops}
 
\author[a]{A.V.~Belitsky,}
\author[b]{L.V.~Bork,}
\author[c]{V.A. Smirnov}
\affiliation[a] {Department of Physics, Arizona State University,  Tempe, AZ 85287-1504, USA}  
\affiliation[b]{Institute for Theoretical and Experimental Physics, 117218 Moscow, Russia}
\affiliation[]{The Center for Fundamental and Applied Research, 127030 Moscow, Russia}
\affiliation[c]{Skobeltsyn Institute of Nuclear Physics, Moscow State University 119992 Moscow, Russia}
\affiliation{Moscow Center for Fundamental and Applied Mathematics 119992 Moscow, Russia} 
  
 \abstract
{In this paper we provide a detailed account of our calculation, briefly reported in arXiv:2209.09263, of a two-particle form factor of the lowest 
components of the stress-tensor multiplet in $\mathcal{N} = 4$ sYM theory on its Coulomb branch, which is interpreted as an off-shell 
kinematical regime. We demonstrate that up to three-loop order, both its infrared-divergent as well as  finite parts do exponentiate in the Sudakov 
regime, with the coefficient accompanying the double logarithm being determined by the octagon anomalous dimension $\Gamma_{\rm oct}$. We 
also observe that up to this order in 't Hooft coupling the logarithm of the Sudakov form factor is identical to twice the logarithm of the null octagon, 
which was introduced within the context of integrability-based computation of four point correlators with infinitely large R-charges. The null octagon 
is known in a closed form for all values of the 't Hooft coupling constant and kinematical parameters. We conjecture that the relation between the 
former and the off-shell Sudakov form factor holds to all loop orders.
}

\begin{document}

\maketitle
\flushbottom
\setcounter{footnote} 0

\section{Introduction}
\label{s1}

The past three decades have witnessed enormous progress in unravelling the structure of S-matrices in gauge theories in various space-time dimensions. 
One playground where this was achieved in a spectacular manner is the maximally supersymmetric Yang-Mills theory in four space-time dimensions, or 
$\mathcal{N}=4$ sYM, for short. The structure of perturbative amplitudes in the latter is qualitatively similar to the one of parton-level scattering in 
Quantum Chromodynamics,--- the theory of strong interaction ---, and it was explicitly employed in aiding cutting-edge multi-loop QCD analyses 
starting from their $\mathcal{N}=4$ sYM counterparts.

A number of analytical frameworks were devised, tested and employed to obtain a plethora of multi-loop and sometimes all-order scattering amplitudes, 
with unitary-based \cite{Dixon:1996wi}, bootstrap \cite{Dixon:2011pw}, duality \cite{Alday:2007hr,Drummond:2007aua,Brandhuber:2007yx} and 
integrability \cite{Basso:2013vsa} methods, to name just a few. The same or similar strategies can be applied to other (dimensionally regularized) 
quantities in $\mathcal{N}=4$ sYM such as form factors and correlation functions. None of these results would be feasible with standard Feynman 
diagrammatic methods, which require calculational power far beyond current computational capabilities.

All-order results for scattering amplitudes \cite{BernBDS,Bern:2006vw} allowed one to explicitly verify theorems regarding infrared (IR) behavior 
and factorization properties of massless amplitudes in gauge theories \cite{Mueller:1979ih,Magnea:1990zb,Sterman:2002qn}. They were found 
in perfect agreement with general considerations stating that IR divergent parts of the color ordered amplitudes have to be given by products of 
two-particle Sudakov form factors \cite{Magnea:1990zb,Sterman:2002qn,BernBDS}. The latter were in turn governed by the ubiquitous cusp
anomalous dimension $\Gamma_{\rm cusp}$ \cite{Polyakov:1980ca,Korchemsky:1987wg}.

These perturbative studies are not of purely academic interest tough, considering the unphysical nature of the model in question, but rather, as we 
briefly touched upon at the very beginning, are of phenomenological relevance as well. For many observables, $\mathcal{N}=4$ results represent 
the ``most complicated'' portion of the ones in QCD and thus can be used to facilitate tedious calculations. This is known under the name of the 
principle of maximal transcendentality \cite{Kotikov:2004er}.

In our present work, a two-particle form factor of a two-scalar field operator in the stress-tensor multiplet will take center stage. Starting with a precocious
two-loop analysis in Ref.\ \cite{vanNeerven:1985ja}, this on-shell observable in the massless $\mathcal{N}=4$ sYM perturbative expansion is currently 
known up to four-loop accuracy \cite{Gehrmann:2011xn,Boels:2017ftb,Huber:2019fxe}. 

In the bulk of the analyses alluded to above, one dealt with the $\mathcal{N}=4$ sYM theory with the exact SU($N$) gauge symmetry. Much less 
attention has been paid to its phase where the latter is broken by non-zero vacuum expectation values (VEVs) of scalar fields present in the model. 
This setup is referred in the literature as to the Coulomb branch \cite{HenrietteCoulomb}. One of the motivations to even address the theory with the 
spontaneously broken gauge symmetry is that so induced particle masses can be regarded as an IR regulator \cite{HennGiggs1,HennGiggs2,HennGiggs3}, 
--- an alternative to the conventional dimensional regularization. As innocent as it may look, the use of the former leaves certain other (space-time) 
symmetries intact, such as the dual conformal symmetry \cite{Drummond:2007aua,Elvang:2013cua}, violated otherwise \cite{HennGiggs1}. Indeed, in 
$\mathcal{N}=4$ sYM with unbroken gauge symmetry, amplitudes and form factors can be made well-defined only in 
$D=4-2\varepsilon$ space-time dimensions and they become singular as $\varepsilon \rightarrow 0$ due to copious emissions of massless states 
and their presence in quantum loops, hence IR divergent. However, in the case of the spontaneously broken gauge symmetry, massive particles will 
play the role of an IR regulator and the limit $\varepsilon \to 0$ can be safely taken from the get-go such that the theory will safely reside in four 
space-time dimensions. IR divergences will now manifest themselves as logarithms of the particle mass $m$ as $m \to 0$. 

The $\mathcal{N}=4$ sYM on the Coulomb branch is also intricately connected to sYM theories in higher dimensions. In this correspondence, massless 
higher dimensional momenta of particles can be interpreted as massive four dimensional ones \cite{HennGiggs1,HenrietteCoulomb,Caron-Huot:2021usw}. 
As a consequence, amplitudes and form factors with massive states in $\mathcal{N}=4$ sYM on the Coulomb branch are expected to be equivalent 
to their counterparts in $\mathcal{N}=1$ sYM in $D=10$ dimensions with loop momenta restricted to the $D=4$ space-time subspace 
\cite{HennGiggs1,HenrietteCoulomb,Caron-Huot:2021usw}. A choice of VEVs can be enforced in such a manner that all external particles are truly 
massless in four dimensions, while nonvanishing masses emerge only for loop states propagating roughly around diagram perimeters. In this massive 
setup, IR factorization properties of scattering amplitudes and Sudakov form factors were briefly discussed in Ref.\ \cite{Henn:2011by}. Results obtained 
there were in line with general expectations about the structure of IR divergences in gauge theories. In particular, the leading IR behavior of form factors 
and amplitudes was controlled by the very same cusp anomalous dimension $\Gamma_{\rm cusp}$, as in the massless case. 

Recently a duality was suggested \cite{Caron-Huot:2021usw}, which relates correlation functions of half-BPS operators with infinitely-large R-charges 
to scattering amplitudes of massive particles, or W-bosons, in planar $\mathcal{N}=4$ sYM on the Coulomb branch in the regime when all states 
which propagate in internal loops are massless. Consistency and gauge invariance of such kinematical regime was advocated by the authors of 
\cite{Caron-Huot:2021usw} using the above massless/massive correspondence between sYM theories in various dimensions. This kinematical 
regime mimics a naive off-shell generalization of a purely massless scattering with unbroken gauge symmetries, so hereafter we will refer to it as 
{\sl off-shell} to distinguish it from the one where massive particles are also present in quantum loops, such as discussed in Refs.\ 
\cite{HennGiggs1,Henn:2011by}. Integrability allowed the authors of Ref.\ \cite{Caron-Huot:2021usw} to obtain a closed-form all-loop expression for 
the scattering amplitude of four W-bosons starting from the four-point correlation function of very heavy half-BPS operators 
\cite{Coronado:2018cxj,Belitsky:2019fan,Belitsky:2020qrm}. This conjecture was supported by a comparative analysis of the ten-dimensional null limit 
of the correlator’s integrand with the $D$-dimensional integrands of four-point amplitudes up to four \cite{Bern:2006ew} and five \cite{Bern:2012uc} loops.

This conjecture allowed the authors of \cite{Caron-Huot:2021usw} to probe the IR behavior of the off-shell four point amplitude to all orders of perturbative 
series and to reveal quite an unexpected result: it turned out that the IR divergences in this case are {\sl not} controlled by the $\Gamma_{\rm cusp}$, as 
previously expected, but rather by a different function of the coupling, the so-called octagon anomalous dimension $\Gamma_{\rm oct}$, which has made its 
debut in the four-dimensional null limit of the the aforementioned four-point large $R$-charge correlator \cite{Coronado:2018cxj,Belitsky:2019fan,Belitsky:2020qrm}. 
Further studies performed in Ref.\ \cite{Bork:2022vat}, this time involving a five-leg off-shell amplitude, supported these results and solidified the role of 
$\Gamma_{\rm oct}$ as the off-shell counterpart of $\Gamma_{\rm cusp}$. These observations immediately raise the question about the structure of  
Sudakov form factors and IR factorization properties of amplitudes in the off-shell kinematical regime. This is a very important and nontrivial endeavor since
these recent findings regarding IR behavior of the off-shell scattering amplitudes are in tension with what was expected previously in $\mathcal{N}=4$ sYM \cite{Drummond:2007aua} as well as in other gauge theories such as QCD \cite{Korchemsky:1988hd,Collins:1989bt}.

The aim of the current paper is to report details of a three-loop computation of the off-shell two-particle form factor and its Sudakov limit in planar 
$\mathcal{N}=4$ sYM. This was first announced in a short note in Ref.\ \cite{Belitsky:2022itf}. Based on this analysis, we confirm that the IR behavior 
of the off-shell Sudakov form factor in $\mathcal{N}=4$ sYM is indeed controlled by $\Gamma_{\rm oct}$ rather than $\Gamma_{\rm cusp}$. Moreover, 
we conjecture a closed-form all-order expression for the finite part of the off-shell Sudakov form factor as well: it is found to be proportional to the 
non-logarithmic ``hard" function of the so-called null octagon $\mathbb{O}_0$, which in turn was introduced within the context of integrability based 
computation of the four point correlation functions with infinitely-large R-charges \cite{Belitsky:2019fan,Belitsky:2020qrm}.

Our subsequent consideration is organized as follows. In Section \ref{s2} we provide a lightening overview of salient facts about IR properties of
amplitudes and form factors in $\mathcal{N}=4$ sYM at different points of the Coulomb branch, its origin and beyond. In Section \ref{s3}, we briefly 
recall the structure of the Sudakov form factor on the Coulomb branch up to two loop order, which was previously known in the literature. We further  
discuss certain assumptions made in our off-shell calculation. In Section \ref{s4}, we present detailed analysis and explicit results for the three-loop 
off-shell form factor and its double logarithmic limit. Based on this computation we also make an all-order conjecture for the off-shell Sudakov form 
factor. In Section \ref{s5}, we summarize our observations regarding IR factorization properties of Coulomb branch amplitudes in the off-shell kinematical 
regime and compare them with statements made in earlier literature. Finally, we summarize. 

\section{IR structure of amplitudes and form factors}
\label{s2}

The IR structure of amplitudes and form factors in gauge theories are closely related to each other. Therefore, before diving into the subject of form 
factors, it is instructive to recall key facts regarding the IR behavior of massless scattering amplitudes in $\mathcal{N}=4$ sYM at the origin of the 
moduli space. This situation is very well understood. 

\subsection{Origin of moduli space}

As in any four-dimensional gauge theories with massless particles, scattering amplitudes in $\mathcal{N}=4$ sYM possess IR divergences. One can 
tame them with conventional dimensional regularization (or its supersymmetric version, dimensional reduction) by considering the theory in 
$D = 4-2\varepsilon$ space-time instead. The IR singularities then manifest themselves as poles in the $\varepsilon$ regulator. For a planar color-ordered 
$n$-particle amplitude $A_n$ with arbitrary helicity content, it is convenient to define the ratio $M_n=A_n/A_n^{\rm tree}$. One can then expect the 
following form of $M_n$ to hold at all-loop orders \cite{Mueller:1979ih,Magnea:1990zb,Sterman:2002qn,BernBDS}:
\begin{align}
\label{LogMDimReg}
\log M_n = - \frac{1}{4} \sum_{i=1}^n\sum_{\ell=1}^{\infty} g^{2 \ell}
\left[
\frac{\Gamma^{(\ell)}_{cusp}}{(\ell \varepsilon)^2}
+
\frac{G^{(\ell)}}{(\ell\varepsilon)}
\right]
\left(\frac{\mu^2}{s_{ii+1}}\right)^{\ell \varepsilon}
+\mathcal{F}_n\big(\{p_i\},a\big)
+O(\varepsilon)
\, .
\end{align}
where the perturbative series is furnished in terms of the 't Hooft coupling $g^2=g_{\rm\scriptscriptstyle YM}^2N/(4\pi)^2$. The generalized Mandelstam 
invariants $s_{ii+1}=(p_i+p_{i+1})^2$ are built from particles' momenta $p_i$ ($i=1, \dots, n$), and $\mu$ is a mass parameter of the dimensional 
regularization. Last but far from being the least, $\Gamma^{(\ell)}_{\rm cusp}$ and $G^{(\ell)}$ are some numerical transcendental coefficients. The function 
$\mathcal{F}_n$ depends on the helicity configuration of the amplitude $A_n$, kinematical invariants as well as the coupling constant, however, it depends
neither on the parameter $\varepsilon$ nor the scale $\mu$. The $1/\varepsilon^2$-pole structure originates from the overlap of soft and collinear 
divergences, where each of them manifests itself individually as $1/\varepsilon$.

We see that IR divergences of the ratio $M_n$ factorize and exponentiate. Their structure is universal for all amplitudes (i.e., independent of particular 
helicity configurations) and is controlled by two sets of coefficients: $\Gamma^{(\ell)}_{\rm cusp}$ and $G^{(\ell)}$. These coefficients in turn define two 
functions of the 't Hooft coupling, the cusp $\Gamma_{\rm cusp}(g)$ and collinear $G(g)$ anomalous dimensions. The first few terms in their
perturbative expansion are
\begin{align}
\label{GCuspAndGPT}
\Gamma_{\rm cusp}(g)
&=
\sum_{\ell=1}^{\infty}\Gamma_{\rm cusp}^{(\ell)} g^{2 \ell}
=
4 g^2 - 8\zeta_2 ^4 + 88 \zeta_4 g^6 + \ldots
\, , \nonumber\\
G(g)
&=
\sum_{\ell=1}^{\infty}G^{(\ell)} g^{2 \ell}
=
-
4 \zeta_3 g^4+ \left( 32 \zeta_5+\frac{80}{3}\zeta_2\zeta_3 \right) g^6 + \ldots
\, .
\end{align}
The above formulas exhibit standard folklore that the leading IR behavior (i.e. $1/\varepsilon^2$ poles) of planar amplitudes in $\mathcal{N}=4$ sYM 
at the origin of moduli space is controlled by $\Gamma_{\rm cusp}(g)$. 

\begin{figure}[t]
\begin{center}
\mbox{
\begin{picture}(0,140)(220,0)
\put(0,80){\insertfig{15}{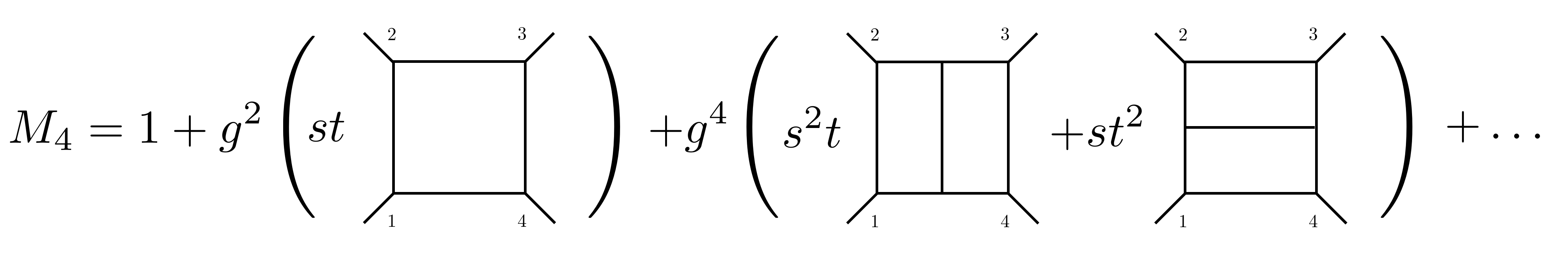}}
\put(0,0){\insertfig{15}{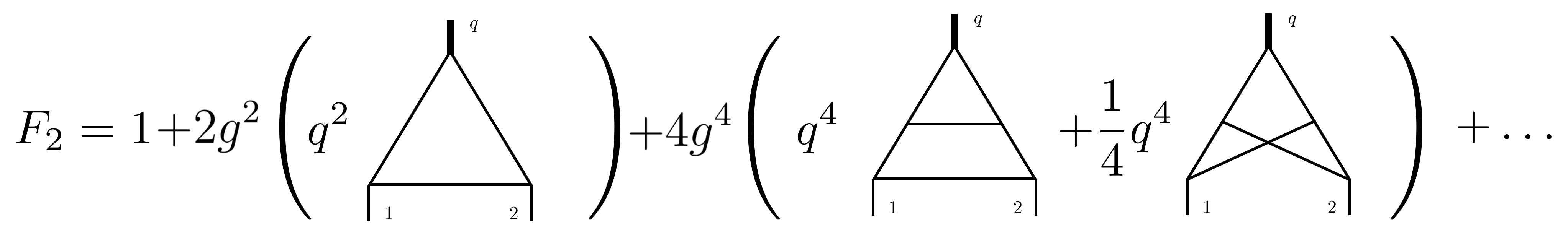}}
\end{picture}
}
\end{center}
\caption{\label{fig1} Two-loop expansion of the four-leg amplitude (top panel) and the two-leg form factor (bottom panel) in terms of scalar integrals. }
\end{figure}

Not only the cusp anomalous dimension, but the whole IR divergent part of the amplitude can be independently defined in terms of matrix elements of some 
local operators. Indeed one can show that the divergent part of the amplitude is given by the product of Sudakov form factors $F_2$, which are determined 
here as matrix elements of an operator from the $\mathcal{N}=4$ sYM stress-tensor supermultiplet and a pair of on-shell states. The lowest component of this 
supermultiplet is given by the operator $\mathcal{O}=\mbox{tr}(\phi_{12}\phi_{12})$, built out of two scalar firelds $\phi_{AB}=\phi_{AB}^a t_a$ from the 
$\mathcal{N}=4$ sYM Lagrangian in the $\bf{6}$ representation of SU$(4)_{\rm R}$ with $t_a$ being the SU($N$) generators in the fundamental representation. 
So one can define $F_2$ as:
\begin{align}
\label{MasslessFFDef}
F_2
= 
\langle0| (\hat{a}^\dagger)^a_{12, p_1} (\hat{a}^\dagger)^a_{12, p_2} \mathcal{O}|0\rangle
/
\langle0| (\hat{a}^\dagger)^a_{12, p_1} (\hat{a}^\dagger)^a_{12, p_2} \mathcal{O}|0\rangle_{\rm tree}
\, .
\end{align}
Here $(\hat{a}^\dagger)^a_{12, p_i}$ is a creation operator of the on-shell scalar $\phi_{12}^a$ with momentum $p_i$, $p_i^2=0$. One can demonstrate 
that $F_2$ will in fact be identical to all other operators from the stress-tensor supermultiplet and all possible pairs of particles from the on-shell 
$\mathcal{N}=4$ sYM supermultiplet \cite{Brandhuber:2011tv,Bork:2011cj}. Information about operator type and particle helicities is encoded in  
$\langle0| (\hat{a}^\dagger)^a_{12, p_1} (\hat{a}^\dagger)^a_{12, p_2} \mathcal{O}|0\rangle_{\rm tree}$, which we factor out. Then one can rewrite 
(\ref{LogMDimReg}) as:
\begin{align}
\label{AmplFactor}
\log M_n = \frac{1}{2}\sum_{i=1}^n\log F_{2}\left(\frac{\mu^2}{s_{ii+1}},g,\varepsilon\right) +\mathcal{F}_n\big(\{p_i\},g\big)-n \, c(g)+O(\varepsilon),
\end{align}
with:
\begin{align}
\label{FF2}
\log F_2\left(\frac{\mu^2}{q^2},g,\varepsilon\right) 
=
-
\frac{1}{2}\sum_{\ell=1}^{\infty} g^{2 \ell}
\left[ \frac{\Gamma^{(\ell)}_{cusp}}{(\ell\varepsilon)^2}
+
\frac{G^{(\ell)}}{(\ell\varepsilon)}
+
c^{(\ell)}\right]
\left(\frac{\mu^2}{q^2}\right)^{\ell\varepsilon}
+
O(\varepsilon).
\end{align}
Here $q=p_1+p_2$ is the off-shell momentum, $q^2 \neq 0$, carried by the operator in question and $c(g)=\sum_\ell c^{(\ell)}g^{2 \ell}$, where $c^{(\ell)}$ 
are some (potentially) scheme-dependent constants. These factorization theorems for $F_2$ and $M_n$ were tested by multiple explicit computations of 
$M_n$ for some values of $n$ and $F_2$ and are in perfect agreement with each other, see, e.g., 
\cite{BernBDS,Bern:2006vw,vanNeerven:1985ja,Brandhuber:2010ad}. 
They are also supported by general theoretical arguments which map IR behavior of amplitudes to UV behavior of cusped Wilson lines 
\cite{Polyakov:1980ca,Korchemsky:1987wg}. Standard renormalization group (RG) machinery can be applied to tackle their UV behavior
\cite{Korchemsky:1988hd,Collins:1989bt}. For example, the all-order structure of $F_2$ (\ref{FF2}) is a result of such analyses 
\cite{Korchemsky:1988hd,Magnea:1990zb,BernBDS}. For illustrative purposes, scalar Feynman integrals contributing to $M_4$ and $F_2$ in the first 
two orders of perturbative series are displayed in Fig.\ \ref{fig1}. It is also worth to mention explicitly that because the operator 
insertion in definition of Sudakov form factor is color singlet, even in the planar limit there will be contributions of non-planar graphs to the Sudakov form 
factor in contrast to the planar nature of amplitude case. This fact makes relations (\ref{AmplFactor}) between amplitudes and form factor especially nontrivial.

In gauge theories with less or without supersymmetry, such as QCD, IR factorization relations similar to (\ref{AmplFactor}) will also hold, but their 
explicit structure will be more involved since one will have to take the running of coupling into consideration
\cite{Korchemsky:1988hd,Collins:1989bt,Magnea:1990zb,Sterman:2002qn}.

\subsection{Coulomb branch}

As we already mentioned in the introductory section, in addition to the coupling constant $g_{\scriptscriptstyle\rm YM}$ and the number of colors $N$, 
the $\mathcal{N}=4$ sYM with SU$(N)$ gauge group possesses another set of free adjustable parameters,--- the VEVs of the six real scalar fields 
$\phi_{AB}$ of the theory, aka moduli. In principle, there are no restrictions on their values and one can consider the theory at any point in its moduli 
space. So what about the above amplitude story away from its origin? As was advocated in \cite{Caron-Huot:2021usw}, it is convenient to use the
aforementioned $D$-dimensional framework together with observations that the integrands of $n=4$ and $n=5$ of planar amplitudes (or rather the
ratio functions $M_n$), have universal structure shared among sYM theories \cite{Boels:2012ie,Mafra:2015mja,Mafra:2008ar}. Then by imposing 
kinematical constraints on the $D$-dimensional integrand, one can obtain integrands in $\mathcal{N}=4$ sYM at a nontrivial position in its moduli space
away from its origin. 

\begin{figure}[t]
\begin{center}
\mbox{
\begin{picture}(0,70)(220,0)
\put(0,0){\insertfig{15}{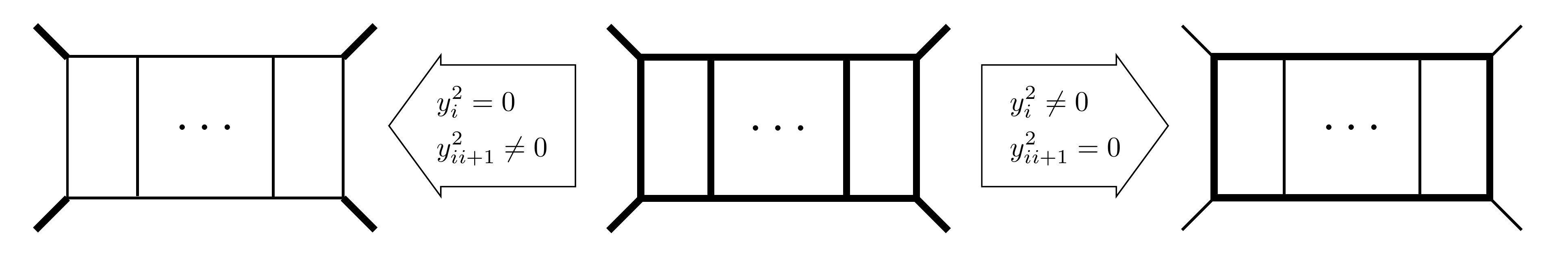}}
\end{picture}
}
\end{center}
\caption{\label{fig2} Various kinematical regimes discussed in the text stemming from a $D$-dimensional progenitor.}
\end{figure}

Ref.\ \cite{Caron-Huot:2021usw} found that it is useful to invoke the $D$-dimensional dual coordinates $X_i$ to impose the aforementioned kinematical 
constraints. These are related to particles' momenta as $p_i = X_{ii+1} \equiv X_i - X_{i+1}$. Namely, all loop integrations $d^DX_l$ are accompanied 
by the constraint $\delta^{D-4}(X_\ell)$. This effectively decomposes all propagator denominators $X^{2}_{i\ell}$ into the four- and extra-dimensional
components, $X^{2}_{i\ell}=x^2_{i\ell}+y_i^2$, where $y_i^2$ must be identified in turn with particle masses $y_i^2\equiv m_i^2$ generated by a specific 
pattern of gauge symmetry breaking. Momenta of external particles encoded in $p_i^2=X_{ii+1}^{2}=0$ should also be decomposed accordingly, 
$X_{ii+1}^{2}=x_{ii+1}^{2}+y_{ii+1}^2$, where once again $D>4$ part is regarded as their mass $y_{ii+1}^2=m_{ii+1}^2$, see Fig.\ \ref{fig2}. The 
specifics of the VEV choice is not relevant for our discussion,--- as long as it is possible ---, so we sweep under the rug these irrelevant details about 
concrete patterns of gauge symmetry breaking, structure of the R-symmetry group after the latter took place etc. All this information is contained in 
$A_n^{\rm tree}$ amplitude, which we factor out anyway. 

As an illustration, let us consider the contribution of the one-loop box integral from Fig.~\ref{fig1}:
\begin{align}
\label{DBoxExample}
\int \frac{d^D X_{\ell} \, \delta^{D-4}(X_{\ell}) \, X_{13}^2 X_{24}^2}{X_{1 \ell}^{2} X_{2 \ell}^{2} X_{3\ell}^{2} X_{4\ell}^{2}}
\mapsto
\int \frac{d^4 x_{\ell} \, X_{13}^{2} X_{24}^2}
{(x_{1\ell}^{2}+y_1^2)(x_{2\ell}^{2}+y_2^2)(x_{3\ell}^{2}+y_3^2)(x_{4\ell}^{2}+y_4^2)}.
\end{align}
There exist two essentially different possibilities. One can consider the situation where all $m_{ii+1}^2=0$. This will correspond to the scattering of 
massless particles via massive states propagating in loops, such as in the four-photon scattering amplitude in QED. Here $y_i^2=m_i^2$ will 
play the role of the IR regulator, and to investigate the IR properties of the amplitude one must evaluate loop integrals in $D=4$ and then approach 
the limit $m_i^2\equiv m^2 \to 0$. For Eq.\ (\ref{DBoxExample}), this will yield
\begin{align}
\label{DBoxExample1}
\int \frac{d^4 x_{\ell} \, x_{13}^{2} x_{24}^2}
{(x_{1\ell}^{2}+m^2)(x_{2\ell}^{2}+m^2)(x_{3\ell}^{2}+m^2)(x_{4\ell}^{2}+m^2)}
\, ,
\end{align}
where everything is four dimensional and $x_{ii+1}^2=p_i^2=0$. In particular, these last conditions imply that $x_{13}^{2}x^{2}_{24}$ is now equal to 
$s_{12} s_{23}$.  

Another interesting option is to consider the opposite limit and put all $y_i^2=m_i^2=0$ first. Then, the external masses $m_{ii+1}^2$ instead will 
play the role of an IR regulator. Once again, it is implemented in such a manner that all integrals are evaluated in $D=4$ and then the limit $m^2_{ii+1} 
\equiv m^2 \to 0$ is taken. Physically, this situation corresponds to the scattering of massive W-bosons in the limit where we neglect all masses of states 
propagating in quantum loops. For our one-loop box (\ref{DBoxExample}), this limit provides:
\begin{align}
\label{DBoxExample2}
\int d^4x_{\ell} \frac{x_{13}^{2}x^{2}_{24}}{x_{1\ell}^{2}x_{2\ell}^{2}x_{3\ell}^{2}x_{4\ell}^{2}},
\end{align}
where every Lorentz invariant is four-dimensional and $x_{ii+1}^2=p_i^2=-m^2 \to 0$. This integral can be re-expressed in terms of the Davydychev-Usyukina 
one loop box function \cite{Usyukina:1992jd,Usyukina:1993ch}. Since the scalar integral families as well as their accompanying numerical coefficients
will be identical to the on-shell case for $n=4,5$, this kinematical regime will be in fact identical to the naive off-shell regularization of purely massless results. 
This was briefly discussed in the earlier literature, see, e.g., \cite{Drummond:2007aua}. At that time such an apparently naive off-shell continuation was obscured 
by potential problems with gauge invariance and, thus, overall consistency of such a procedure. Relations to higher dimensional sYM theories were not 
(widely) known or explored back then. 
 
Let us consider the situation with the scattering of massless external particles via massive virtual states first. Using the aforementioned prescription, 
explicit $n=4$ three loop and $n=5$ two loop computations were performed\footnote{At two-loop level, they were also verified by explicit Feynman 
diagram calculations \cite{HennGiggs1}.} \cite{HennGiggs1,HennGiggs2,HennGiggs3}. Results of these computations allow one to conjecture the 
following IR factorization formula for the ratio function $M_n$:
\begin{align}
\label{LogMHigssg}
\log M_n = - \frac{1}{8}\sum_{i=1}^n\left[ \Gamma_{\rm cusp}(g) \log^2\left( \frac{m^2}{s_{ii+1}} \right)
+ 
2
\widetilde{G}(g)\log\left( \frac{m^2}{s_{ii+1}} \right)\right]
+
\mathcal{F}_n\big(\{p_i\},g\big)+O(m^2)
\, .
\end{align}
Here $\widetilde{G}(g)$ may potentially be different from the pure massless case (\ref{LogMDimReg}). Based on the results of Refs.\
\cite{HennGiggs1,HennGiggs2,HennGiggs3}, one can also expect that the hard function $\mathcal{F}_n$ here is identical to the purely massless case. 
We see that this situation is essentially equivalent to the massless case (i.e., the theory at the origin of its moduli space) with the replacement 
of $1/\varepsilon$ poles with $\log m^2$. Leading IR logarithms are still controlled by $\Gamma_{\rm cusp}(g)$, which is in line with the folklore that 
$\Gamma_{\rm cusp}(g)$ is ``the ultimate IR anomalous dimension" and all IR limits of the theory should be controlled by $\Gamma_{\rm cusp}(g)$ 
of that theory. It is also curious to mention for Eq.\ (\ref{LogMHigssg}) to hold, it is sufficient to retain nonvanishing $m_i^2$ only in propagators which 
form a closed frame around graph sites. All other masses can be considered strictly set to zero, see, e.g., the right-hand side of Fig.\ \ref{fig2} for the 
$n=4$ amplitude.

The opposite situation of the massive particle scattering via massless virtual states revealed, however, a different picture. Based on the three loop 
$n=4$ and two loop $n=5$ computations \cite{Caron-Huot:2021usw,Bork:2022vat}, one can conjecture the following IR factorization formula:
\begin{align}
\label{M5div}
\log M_n= - \frac{1}{4} \sum_{i=1}^n \Gamma_{\rm oct}(g) \log^2\left( \frac{m^2}{s_{ii+1}} \right)
+
\widetilde{\mathcal{F}}_n\big(\{p_i\},g\big)
+
O(m)
\, ,
\end{align}
with $\Gamma_{\rm oct}(g)$ being a different function of 't Hooft coupling compared to $\Gamma_{\rm cusp}(g)$:
\begin{align}
\Gamma_{\rm oct}(g)=4 g^2 - 16 \zeta_2 g^4 + 256 \zeta_4 g^6 + \ldots
\, ,
\end{align}
and (potentially) different $\widetilde{\mathcal{F}}_n$ compared to pure massless case. More accurately for $n=4,5$ examples the kinematical dependence 
of $\widetilde{\mathcal{F}}_n$ and $\mathcal{F}_n$ was identical, but the dependence on coupling constant $g$ was different, which can be captured by 
$\Gamma_{\rm oct} \mapsto \Gamma_{\rm cusp}$ replacement.

This unexpected result immediately raises the question about the IR factorization properties of amplitudes in the off-shell kinematical regime, i.e., can 
their IR divergent parts be captured by the product of the off-shell Sudakov form factors? The ultimate goal of this article is to shed light on these questions 
and we will address them in details in the next sections.

\section{Form factors in $\mathcal{N}=4$ sYM}
\label{s3}

Before discussing the off-shell regime for production of external (massive) particles by the operator $\mathcal{O}$ from the vacuum, let us briefly discuss 
the opposite situation for scattering of massless particles via massive virtual states. 

\subsection{Coulomb branch: massive internal lines}
\label{s3a}

For simplicity ,we will choose a single mass $y_i^2=m^2 \ll 1$ for all $i$ and then take the limit $m^2 \to 0$. In the pure massless case with the unbroken 
gauge symmetry, the Sudakov form factor $F_2$ reads up to two loops 
\begin{align}
\label{LogMHigssgInt1}
F_2=1+g^2 F_2^{(1)}+g^4 F_2^{(2)}
+
\ldots \, .
\end{align}
in terms of the following set of scalar integrals, shown graphically in Fig.\ \ref{fig1},
\begin{align}
\label{LogMHigssgInt2}
F_2^{(1)}&=2Q^2 \, T_{1,1}^{\text{on-shell}},
\nonumber\\
F_2^{(2)}&=4Q^4 \left[ T_{2,1}^{\text{on-shell}}+\frac{1}{4}~T_{2,2}^{\text{on-shell}}\right],
\end{align}
where we introduced Euclidean momentum transfer $Q^2 = - q^2 > 0$. These results for $p_i^2=0$ were (re)derived by different methods by multiple 
authors \cite{vanNeerven:1985ja,Bork:2010wf,Gehrmann:2011xn}. As in the amplitude case, one can argue that the integrands in these expansionx are 
identical among sYM theories in all $D$'s \cite{Gehrmann:2011xn}. So it is tempting to use the massless/massive prescription of \cite{Caron-Huot:2021usw} 
to obtain  scalar integral representation for form factors as well starting from Eq.\ (\ref{LogMHigssgInt2}). One faces an immediate obstacle, however, due to 
the presence of non-planar\footnote{These graphs are endowed nevertheless by leading color structures due to the fact that the operator vertex
$\mathcal{O}$  is a unity matrix in the SU$(N)$ group space contrary to other external legs which are in its adjoint representation.} graphs, e.g., 
$T_{2,2}^{\text{on-shell}}$ is the first example of such scalar integral. A general consensus is that there is no well-defined way to introduce dual 
coordinates for such integrals. Despite this fact, one can still insert masses in massless propagators in non-planar $D=4$ dimensional integrals 
``by hand" akin to how it is done for planar integrals, assuming that there is a way to choose $D$-dimensional momenta to replicate these mass 
insertions. Hereafter, we will adopt exactly this hands-on approach for obtaining candidates integrals for massive integrands from purely massless ones.
 
\begin{figure}[t]
\begin{center}
\mbox{
\begin{picture}(0,60)(240,0)
\put(0,0){\insertfig{17}{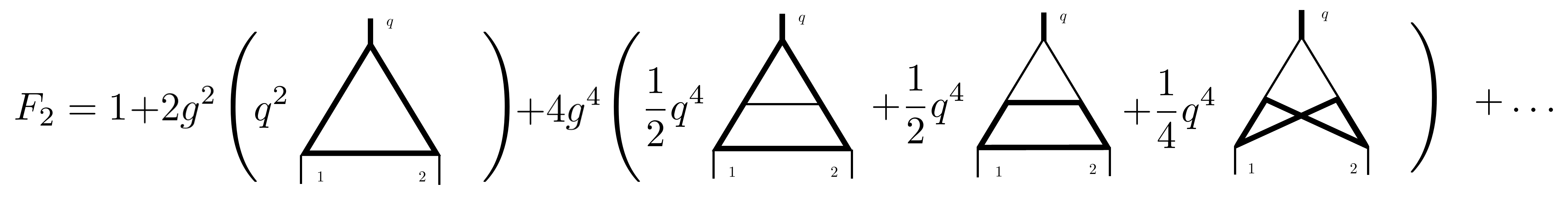}}
\end{picture}
}
\end{center}
\caption{\label{fig3} Scalar integrals determining the second order perturbative expansion of the on-shell form factor regularized by massive 
internal lines.}
\end{figure}
 
In current case, there are two possible ways to insert massive propagators to form closed frames in the $T_{2,1}^{\text{on-shell}}$ integral, which 
will lead to finite four-dimensional results. It is natural to consider both of them with equal apportions, i.e.,  $\frac{1}{2}$ coefficients, see Fig.\ 
\ref{fig3}. The choice of the massive frame in the nonplanar integral $T_{2,2}^{\text{on-shell}}$, which make this integral finite in $D=4$, is unique.  
This results in the following conjectured expansion for the on-shell form factor with massive internal loops:
\begin{align}
\label{LogMHigssgInt}
F_2^{(1)}
&=
2Q^2 \, T_{1,1}^{\text{massive}},\nonumber\\
F_2^{(2)}
&= 
4Q^4 \left[ \frac{1}{2} T_{2,1}^{\text{massive},a}+\frac{1}{2} T_{2,1}^{\text{massive},b}+\frac{1}{4} T_{2,2}^{\text{massive}} \right],
\end{align}
where, as was explained earlier, $T_{i,j}^{\text{massive}}$ are scalar integrals corresponding to the very same graphs as $T_{i,j}^{\text{on-shell}}$ 
in (\ref{LogMHigssgInt2}), but massive instead of massless propagators, see Fig.\ \ref{fig3}. This decomposition can in fact be derived \cite{Henn:2011by} 
standard perturbation theory in $\mathcal{N}=4$ sYM where a specific pattern of spontaneous gauge symmetry breaking is chosen, see Refs.\ 
\cite{HennGiggs1,HennGiggs2,HennGiggs3} for details. This provides a solid endorsement for our approach of uplifting massless contributions
off of their mass shell.

The small-mass expansion for all $T_{i,j}^{\text{massive}}$ integrals leads to the following result \cite{Henn:2011by}:
\begin{align}
\label{LogMHigssg4}
\log F_2\left(\frac{- m^2}{Q^2},g\right) 
= 
-
\frac{1}{4}\Gamma_{\rm cusp} (g) \log^2\left( \frac{- m^2}{Q^2} \right)
-
\widetilde{G}(g)\log\left( \frac{- m^2}{Q^2} \right)
+
\widetilde{c}(g)
+
O(m^2)
\, .
\end{align}
Here as before $\Gamma_{\rm cusp}(a)$ and $\widetilde{G}(a)$ identical to those of (\ref{GCuspAndGPT}) at this order of perturbative series, i.e., 
two loops. 

One can expect that the structure of (\ref{LogMHigssg}) will hold to all orders in $g$ with the same $\Gamma_{\rm cusp}(g)$ but (potentially) 
different $\widetilde{G}(g)$ and $\widetilde{c}(g)$ compared to Eq.\ (\ref{FF2}). This means that the factorization formula on this massive
Coulomb branch will be identical to the one of the massless case (\ref{AmplFactor}). The structure of the Sudakov form factor itself is also very 
similar to purely massless case with bold replacements of all poles $1/\varepsilon$ by logarithms $\log m^2$.

The Sudakov form factor satisfies an evolution equation \cite{Korchemsky:1988hd,Collins:1989bt,Drummond:2007aua}, which in the  massless 
case is given by
\begin{align}
\label{EvolEq}
\left( \frac{\partial}{\partial \log \mu^2} \right)^2 \log F_2 = - \frac{1}{2} \Gamma_{\rm cusp}(g)
\, .
\end{align}
We see from (\ref{LogMHigssg4}) that in the case of the massive kinematical regime, this evolution equation will be intact provided on replaces 
$\log \mu^2 \mapsto \log m^2$. At the time of its derivation, this result was completely in line with general expectations that all IR 
physics in gauge theories under consideration must be controlled by $\Gamma_{\rm cusp}$ of that theory and that $\Gamma_{\rm cusp}$ is 
the ``ultimate anomalous dimension" of IR physics. The appearance of $\Gamma_{\rm cusp}$ in the form factors and amplitudes in this massive 
kinematical regime was considered for granted and was interpreted as the consequence of scheme independence of $\Gamma_{\rm cusp}$ 
\cite{HennGiggs1}.

\subsection{Coulomb branch: off-shell  regime}
\label{s3b}

\begin{figure}[t]
\begin{center}
\mbox{
\begin{picture}(0,170)(190,0)
\put(0,0){\insertfig{13}{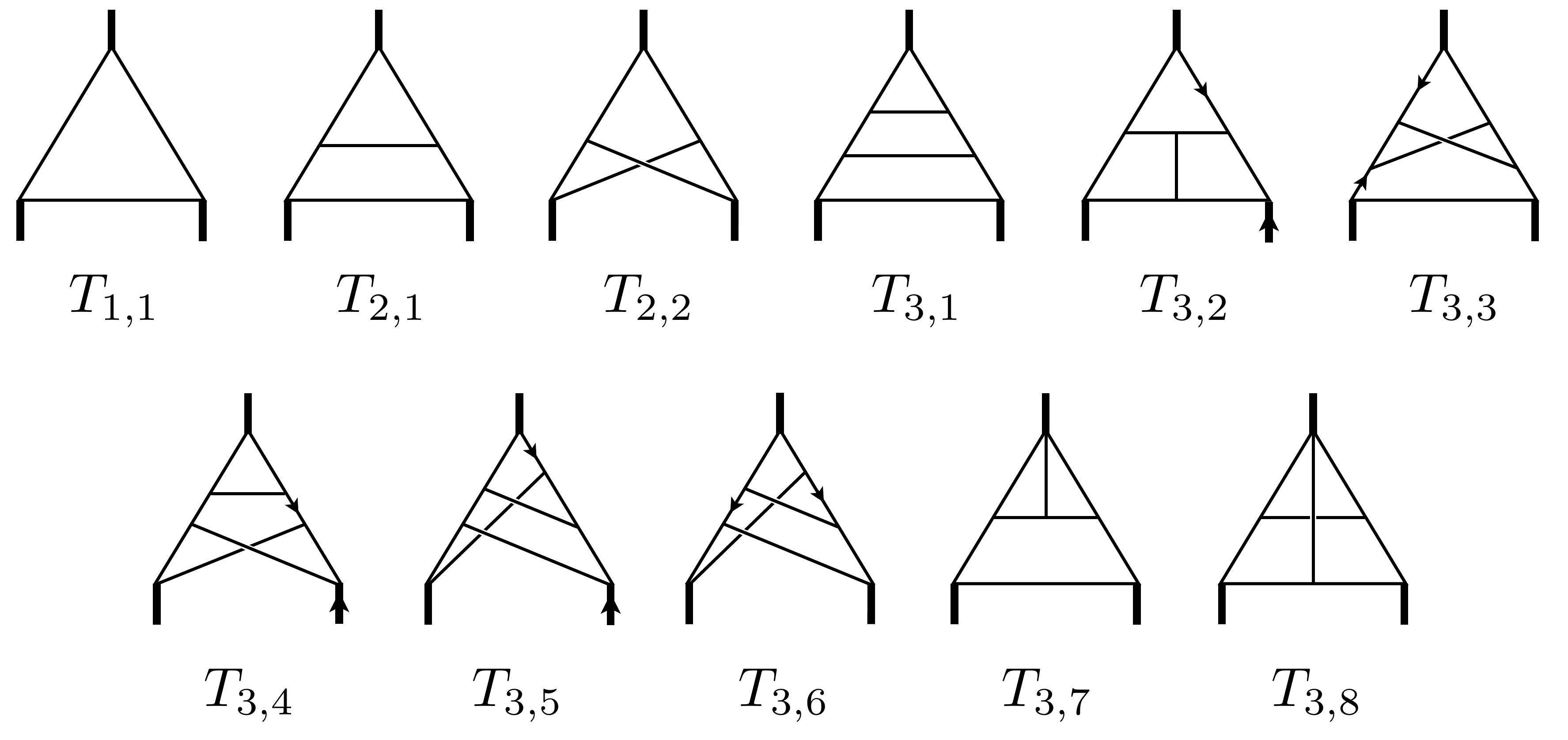}}
\end{picture}
}
\end{center}
\caption{\label{fig4} Scalar integrals contributing to the three-loop off-shell form factor. Arrows on the lines corresponds to the presence of the numerator $(p_a+p_b)^2$, where $p_a$ and $p_b$ are the momenta flowing through corresponding lines.}
\end{figure}
 
Let us now return to the main observable of our interest, i.e., the Sudakov form factor in the off-shell kinematical regime. Based on the two-loop 
purely massless result (\ref{LogMHigssgInt}) and making use of the $D$-dimensional approach described in the previous sections, we can 
conjecture that the perturbative expansion of the Sudakov form factor of W-bosons
\begin{align}
\label{LogMoffshell1}
F_2 = 1 + g^2 F_2^{(1)} + g^4 F_2^{(2)} + g^6 F_2^{(3)}
+
\ldots\, ,
\end{align}
is determined up to two loop order by the following scalar integrals 
\begin{align}
\label{LogMoffshell2}
F_2^{(1)}
&=
2Q^2 T_{1,1} \, , \nonumber\\ 
F_2^{(2)}
&=
4Q^4 \left[ T_{2,1}+\frac{1}{4} T_{2,2}\right] 
\, ,
\end{align}
with very same graphs representation as in (\ref{LogMHigssgInt2}), but now with momenta of external particles being $p_i$ massive, or off-shell.
On the other hand, all internal lines (propagators) are massless, see Fig.\ \ref{fig4}. Note that we currently choose $m^2=-p_i^2$ in contrast to the 
previous section, where $m$ was the mass parameter in the massive propagators $1/(k^2-m^2)$.

In the following, we will use the notation\footnote{We hope that there will be no confusion with a Mandelstam invariant for the four-point amplitude.} 
$t \equiv m^2/Q^2>0$ for the dimensionless ratio entering all integrals.  The one- and two-loop integrals $T_{1,1}$ and $T_{2,1}$, $T_{2,2}$ can be 
expressed in terms of the Davydychev-Usyukina $\ell$-loop box ladder functions $\Phi_{\ell}(x,y)$ \cite{Usyukina:1992jd,Usyukina:1993ch,Usyukina:1994iw} 
as follows
\begin{align}
\label{12loopIntegralsPlanar}
Q^2 T_{1,1}
&=
- \Phi_1(t,t) \, , \qquad 
Q^4 T_{2,1} = \Phi_2(t,t)
\, , \\
\label{2loopIntegralNPlanar}
Q^4 T_{2,2}
&=
\left[\Phi_1(t,t)\right]^2.
\end{align}
where we grouped them together according to their (non)planarity. In turn, the Davydychev-Usyukina functions $\Phi_{\ell}(x,y)$ can be solely written 
in terms of (poly)logarithms,
\begin{align}
\label{BoxFunct}
\Phi_\ell(x,y)=-\sum_{j=\ell}^{2\ell}\frac{j!(-1)^{j}\log^{2 \ell - j}\left( \frac{y}{x} \right)}{\ell! (j - \ell)! (2 \ell - j)!}
\frac{\mbox{Li}_{j} \big(-(\rho x)^{-1}\big) - \mbox{Li}_{j}\big( -(\rho y)^{+1}\big)}
{\lambda} 
\, ,
\end{align}
with $\rho$ and $\lambda$ being functions of $x$ and $y$,
\begin{align}
\lambda(x,y)=[(1-x-y)^2-4xy]^{1/2}\, , \qquad \rho(x,y)=2 [1-x-y-\lambda(x,y)]^{-1}.
\end{align}

It is worth to realie that Eq.\ (\ref{2loopIntegralNPlanar}) is highly non-trivial since it re-expresses the nonplanar integral $T_{2,2}$  in terms of the 
planar $T_{1,1}$. It is well-known that all planar scalar integrals\footnote{The integral $T_{2,1}$ can be obtained from the double box integral using a 
limiting procedure from Refs.\ \cite{Usyukina:1992jd,Usyukina:1993ch}. The same is applicable to other planar integrals in Eq.\ (\ref{LogMoffshell1}),
at least up to three loops.}, including $T_{2,1}$ above, are invariant\footnote{Of course, these have to be accompanied by certain prefactors built 
out from external particle momenta to render them dimensionless.} with respect to the so-called dual conformal symmetry, i.e., conformal
boosts in the momentum space \cite{Drummond:2007aua}, which is a harbinger of integrability of $\mathcal{N}=4$ sYM. We will see in the 
next section that relations akin (\ref{2loopIntegralNPlanar}) between non-planar and planar integrals is likely to be a general pattern of the Sudakov form 
factor in $\mathcal{N}=4$ in the off-shell kinematical regime.

For the first two $\ell$'s, i.e., $\ell=1,2$, the small-$t$ expansion of the Davydychev-Usyukina functions immediately reads
\begin{align}
\label{BoxSmallMassExp1}
\Phi_1(t,t)&=\log^2t+2\zeta_2+O(m^2),\nonumber\\
\Phi_2(t,t)&=\frac{\log^4t}{4}+3\zeta_2\log^2t+\frac{21\zeta_4}{2}+O(m^2) \, .
\end{align}
Let us direct the reader's attention to the fact that contrary to the case of the dimensional regularization, used in the purely massless cases, 
there is no analog of $\varepsilon \times 1/\varepsilon$-interference between different orders in coupling constant $g$, and relation like
Eq.\ (\ref{BoxSmallMassExp1}) are sufficient to completely determine $\log F_2 $ up to terms $O(m^2)$ and three loop accuracy. At this
point, let us quote the expansion for $\Phi_3 (t,t)$ which will be useful in what follows
\begin{align}
\label{BoxSmallMassExp3}
\Phi_3(t,t)=\frac{\log^6 t}{36}+\frac{5\zeta_2}{6}\log^4 t
+\frac{35\zeta_4}{2}\log^2 t+\frac{155\zeta_6}{4}
+
O(m^2)
\, .
\end{align}
Substituting these in $\log F_2$ and expanding, in turn, $\log F_2$ in powers of $g$ up to two-loop accuracy, we obtain \cite{Bork:2022vat}:
\begin{align}
\log F_2\left(t,g\right) 
= 
\left[ -2 g^2 + 8 \zeta_2 g^4 + \ldots \right] \log^2 t
+
\left[ - 4 \zeta_2 g^2 + 32 \zeta_4 g^4 + \ldots\right]
+
O(m^2)
\, .
\end{align}
The structure of this result is in line with general expectations about exponentiation of IR logarithms, however, there are some important differences. 
Indeed we see that IR logarithms exponentiate but the coefficient accompanyin $\log^2 m^2$ is different from $-\Gamma_{\rm cusp}(g)$ divided by $4$. 
Note also that the analogue of the collinear anomalous dimensions $G(g)$, that is $\widetilde{G}(g)$, is completely missing in this case, compared to 
(\ref{LogMHigssg}). A naked eye inspection of the coefficient accompanying the double logarithm as well as the finite piece \cite{Bork:2022vat}, 
allows one to verify that these are in agreement with the leading two terms of the perturbative expansion of the null octagon encoded by the two functions 
of 't Hooft coupling, which are known exactly in terms of elementary functions \cite{Belitsky:2019fan,Belitsky:2020qrm,Belitsky:2020qzm}
\begin{align}
\label{ExactGammaOct}
\Gamma_{\rm oct}(g) 
&= \frac{2}{\pi^2}\log \left[ \cosh \left(2 \pi g \right)\right]=-4 g^2 + 16 \zeta_2 g^4 
+ \ldots
\, , \\
\label{ExactDet}
D(g)
&
=
\frac{1}{4}\log \left[ \frac{\sinh( 4 \pi g )}{4\pi g} \right] = 4 \zeta_2 g^2 - 32 \zeta_4 g^4 
+ \ldots
\, .
\end{align}

The above off-shell kinematical regime was discussed in early literature, see, e.g., \cite{Korchemsky:1988hd,Drummond:2007aua}. It was 
predicted there that the coefficient before $\log^2 m^2$ should merely be given by $-\Gamma_{\rm cusp}$, that is, twice larger compared to the
purely massless case (\ref{FF2}) and (\ref{EvolEq1}). At one loop this is indeed the case, and the origin of the doubling is very well understood
\cite{Sudakov:1954sw,Jackiw:1968zz,Fishbane:1971jz,Mueller:1981sg}. The latter is a consequence of an additional integration domain over the 
loop momentum in $T_{1,1}$ integral, dubbed the ultra-soft regime, compared to the on-shell case. This domain provides leading contribution on 
par with the soft-collinear regions intrinsic to both on- and off-shell integrals. However, as we observe, at higher loop orders this simple doubling 
relation is no longer true. Thus, a calculation of the three loop correction is highly desirable to clarify this clash and to support the claim that the 
coefficient of $\log^2 m^2$ is indeed proportional to $\Gamma_{\rm oct}$ rather that $\Gamma_{\rm cusp}$. This is what we are set to demonstrate 
in the next section.

\section{Off-shell form factor at three loops and beyond}
\label{s4}

In this section, we present the main result of the current work.

\subsection{Integral basis at three loops}\label{s4a}

The massless form factor at three loop order was first evaluated in \cite{Gehrmann:2011xn}. We employ their basis of scalar integrals and 
cast this contribution $F_2^{(3)}$ to Eq.\ (\ref{LogMoffshell1}) into the form
\begin{align}
\label{LogMoffshell3}
F_2^{(3)}
=
8Q^4 
\left[
Q^2 \, T_{3,1}-\frac{1}{4}T_{3,2}+\frac{1}{2}T_{3,3}+\frac{1}{2}T_{3,4}-\frac{1}{2}T_{3,5}
-\frac{1}{2}T_{3,6}-\frac{1}{2}T_{3,7}+\frac{1}{4}T_{3,8}
\right]
\, ,
\end{align}
see Fig.\ \ref{fig4} for graphical representation of individual $T_{3,i}$'s. Our goal is to evaluate $T_{3,i}$. More precisely, we are interested in their 
small-$m$ behavior up to $O(m^2)$. This was accomplished making use of two independent methods. We will discuss them in turn.

\subsection{Evaluating $T_{3,i}$ integrals with differential equations}
\label{s4b}

The first approach that we used is based on solving differential equations~\cite{Kotikov:1990kg,Gehrmann:1999as,Henn:2013pwa} for these 
integrals at general values of $Q^2$ and $m^2$, with $p_1^2 = p_2^2=-m^2$. In this manner, we obtained three-loop corrections in exact
kinematics, i.e., not just the limit $m \to 0$. In fact we adopted an earlier analysis from Ref.~\cite{Pikelner:2021goo} where these integrals
were calculated at the symmetrical point, $p_1^2 = p_2^2 = q^2 = -1$. The homogeneity of $T_{3,i}$ in the two kinematical invariants allows
us to factor out their mass dimension in terms of $Q^2$ and thus focus solely on their dependence on the dimensionless ratio $t = m^2/Q^2$.

The first step on the way to find the so-called canonical form of the differential equations \cite{Henn:2013pwa} 
\begin{align}
\partial \bit{J} = \varepsilon \bit{A} \cdot \bit{J}
\, ,
\end{align}
is to perform an IBP reduction for all $T_{3,i}$ in order to reveal the initial set of master integrals $\bit{I}$ involved in their representation. This can be 
achieved by any available for this purpose software. We relied on {\tt FIRE6} \cite{Smirnov:2019qkx} and {\tt LiteRed} \cite{Lee:2012cn}. For the 
resulting basis of preliminary master integrals, we then constructed a set of differential equation in the $t$-variable. To transform the resulting differential 
equations to a Fuchsian (aka  {\tt dLog}) and $\varepsilon$-form, it is convenient to introduce a new variable by $t = -(z-1)^2/z$. This choice 
rationalizes potential square roots in the matrix $\bit{A}$. Then the singular points $t=\{1,0,\infty,4\}$ of the latter are transformed into 
$z = \{\exp (\frac{i \pi}{3}),1,0,-1\}$. Let us emphasize that the appearance of the sixth root of unity $\sigma = \exp (\frac{i \pi}{3})$ in this problem is a feature 
that shows up for the first time at the three-loop order. 

Making use of the public codes {\tt CANONICA} \cite{Meyer:2017joq} and {\tt Libra} \cite{Lee:2014ioa,Lee:2020zfb}, we transformed the initial system 
of differential equations for the integrals $\bit{I} = \bit{T} \cdot \bit{J}$ first to the {\tt dLog} and ultimately to the $\varepsilon$-form, with the $\bit{A}$ 
matrix having only Fuchsian singularities
\begin{align}
\label{eq:deMatEpForm}
\bit{A}
=
\frac{\bit{R}_{0}}{z}
+ \frac{\bit{R}_{1}}{z-1}
+ \frac{\bit{R}_{-1}}{z+1}
+ \frac{\bit{R}_{\sigma}}{z- \sigma}
+ \frac{\bit{R}_{\sigma^*}}{z-\sigma^\ast}
\, .
\end{align}

The main advantage of the canonical form of the resulting differential equations is that their general solution
\begin{align}
\bit{J}
=
P \exp \left( \varepsilon \int dz \bit{A} \right)
\cdot
\bit{J}_0
\, ,
\end{align}
when expanded in the power series in $\varepsilon$ can be easily integrated order-by-order up to the desired weight-six contributions.
The results are naturally expressed in terms of multiple Goncharov polylogarithms~\cite{Goncharov:2001iea} or, presently, via harmonic
polylogarithms (HPL) \cite{Remiddi:1999ew}. The integration constants $\bit{J}_0$ were fixed with the help of boundary conditions in 
the limit where one of the two external momenta was considered large in the Euclidean sense. In this limit  well-known graph-theoretical 
methods can be used to perform systematic expansions. These prescriptions are implemented in the computer codes {\tt EXP}
\cite{Harlander:1998cmq,Seidensticker:1999bb} and {\tt MINCER}~\cite{Gorishnii:1989gt,Larin:1991fz}, which we relied on.

As a result of our analysis, we deduced the analytical form for all integrals $T_{3,i}$ required for three-loop off-shell form factor. For example, 
we obtained for $T_{3,7}$ 
\begin{align}
Q^4 T_{3,7}
&=
-\frac{40 \, \text{H}_{6}}{(t-1)^3 (t+1) t^2}
+
\frac{4 \pi ^4 \, \text{H}_{0,0}}{9 (t-1)^3 (t+1) t^2}
+
\frac{20 \pi ^2 \, \text{H}_{0,0,0,0}}{3 (t-1)^3 (t+1) t^2}
\nonumber\\
&+
\frac{ 20\pi i \, \text{H}_{0,0,0,0,0}}{(t-1)^3 (t+1) t^2}
-
\frac{20 \, \text{H}_{0,0,0,0,0,0}}{(t-1)^3 (t+1) t^2}
+
\frac{8 \pi ^6 }{189 (t-1)^3 (t+1) t^2}
\, ,
\end{align}
where we used the conventional nomenclature
\begin{align}
\text{H}_{a_1,\ldots,a_n}\equiv \text{HPL}_{a_1,\ldots,a_n}(t)
\, .
\end{align}
The rest of integrals are relegated to the attached {\tt Mathematica} notebook {\tt 3loopTs.nb}. Eventually, these were expanded in the limit $t \to 0$. 
Since all contributions are expressed in terms of HPL, this is a rather straightforward task and can be systematically performed by means of the 
so-called shuffle relations. For instance, for $T_{3,7}$ we find\footnote{If the appropriate branch of HPL
functions is chosen.}
\begin{align}
Q^4
T_{3,7}
=
\frac{\log^6 t}{36}+\frac{5\zeta_2}{6}\log^4 t +\frac{35\zeta_4}{2}\log^2 t+\frac{155\zeta_6}{4}+O(m^2)
\, ,
\end{align}
It is curious to observe that this turns out to be the small-$t$ expansion of the Davydychev-Usyukina function $\Phi_3(t,t)$ in disguise. Explicit form
of these series for all other integrals $T_{3,i}$ can be found in Appendix \ref{a2}.

\subsection{Evaluating $T_{3,i}$ integrals with expansion by regions}
\label{s4c}

In order to cross check our findings, we relied on yet another method to evaluate the small-$t$ expansion of the three-loop integrals. It is based on 
a strategy of the expansion by regions \cite{Beneke:1997zp} (see also \cite{Smirnov:2002pj,Smirnov:2012gma}). It was originally introduced to tackle
threshold expansions of Feynman integrals \cite{Beneke:1997zp} and later generalized to any limit. The limit under consideration in the present 
work, i.e., $t \to 0$, is intrinsic to Minkowski space-time and cannot be formulated in Euclidean kinematics\footnote{Limits of Feynman integrals typical 
to Euclidean space, such as a off-shell large momentum expansion, receive support from the Wilson operator product expansion. The latter can
be formulated in a graph-theoretical language as in Refs.\ \cite{Chetyrkin:1988zz,Chetyrkin:1988cu,Gorishnii:1989dd,Smirnov:1990rz}, see 
\cite{Smirnov:1994tg,Smirnov:2002pj} for comprehensive reviews.}. The essence of the expansion by region consists in classification of loop-momentum
integrands  with regard to their scaling in a small parameter involved, i.e., $t$ for the case at hand. The contributions of these so-called regions are then evaluated
according to the instructions formulated in Ref.\ \cite{Beneke:1997zp} by extending loop integrals to the {\sl entire} infinite space without any kinematical 
restrictions. Setting to zero all emerging scaleless integrals, one obtains desired asymptotic expansion by summing up non-vanishing contributions of the
regions.

Remarkably, this fairly dubious procedure formulated in the momentum space works extremely well in practice. However, insisting on the 
momentum-space language it proves rather difficult to reveal all regions in a given limit following the decomposition of loop momenta in terms of 
hard, collinear, soft and ultrasoft. It is the use of the Feynman parametric representation, see Eq.\ (\ref{Fp}) below, 
which allowed one to alleviate this drawback and provided the possibility to develop a systematic algorithm \cite{Smirnov:1999bza,Pak:2010pt,Jantzen:2012mw} 
and, moreover, to implement it in the computer code {\tt asy} \cite{Pak:2010pt,Jantzen:2012mw}. Within this algorithm, relevant regions correspond to facets 
(i.e., faces of maximal dimension) of a Newton polytope connected with two Symanzik polynomials in the Feynman parametric representation.
In fact, the expansion by regions can be applied with the use of the code {\tt asy} to any parametric integrals over\footnote{For other domains, one should
first map it to $R_+^N$ and then proceed with {\tt asy}. An example of its application to integrals, which are not Feynman integrals, can be found in 
\cite{Belitsky:2021huz} where the initial integration domain was a multidimensional unit cube.} $R_+^N$ of products of polynomials raised to powers linearly 
depending on the regularization parameter $\varepsilon$. Expansion by regions has up to now the status of experimental mathematics. However, to date
there are no known examples where it fails. Let us refer to \cite{Semenova:2018cwy,Smirnov:2021dkb} for discussions of possible ways to prove this strategy.

For Feynman integrals with integrands determined by a product of $N$ propagators $1/(p^2 - m^2 + i 0)^{a_i}$, the corresponding Feynman parametric 
representation is an integral over a projective $R_+^N$,
\begin{align}
\label{Fp}
I_{a_1, \dots, a_N} =  \frac{\left(i \pi^{D/2} \right)^L \Gamma(a-L D/2)}{\prod_i\Gamma(a_i)}
\int_0^\infty
\prod_{i=1}^N x_i^{a_i-1} {\rm d} x_i
\delta\left( \sum x_i-1\right)
\, 
U^{a-(L+1) D/2} F^{L D/2 -a}
\, ,
\end{align}
where $a=\sum a_i$, $L$ is the number of loops and $F=U\sum m^2_i x_i-V$. The functions $U$ and $V$ are the two Symanzik polynomials given by 
the well-known formulas with summations over graph's trees and 2-trees, respectively (see, e.g., \cite{Smirnov:2012gma}). To apply {\tt asy} to (\ref{Fp}), 
it proves more convenient to exploit it as a part of the {\tt FIESTA5} distribution package \cite{Smirnov:2013eza,Smirnov:2021rhf}.  The advantage of its 
use is that the command {\tt UF} will generate $U$ and $V$ automatically as well. In fact, the folklore Cheng-Wu theorem allows one to choose the sum 
over Feynman parameters in the argument of the delta-function to be over any nonempty subset of indices. For example, one can take 
$\delta\left(x_{i_0}-1\right)$ for a conveniently chosen $i_0$. In circumstances when some indices $a_i$ are negative, i.e., it's a numerator, the 
corresponding parametric integral is obtained by a limiting procedure, with a result which has no integration over the corresponding parameter and 
involves extra polynomials in the integrand. So, in Eq.\ (\ref{Fp}) we implied that all propagator indices $a_i$ are positive and, if some indices 
are negative {\tt FIESTA5} immediately yields corresponding expressions as well.

The heuristic formulation of the expansion by regions in the momentum space alluded to above can be repeated to the letter for parametric integrals as 
well. Let us emphasize, however, that the notion of a region here literally implies certain scaling of integration variables with powers of the small parameter 
in the problem. Our goal is then to select scalings that generate, after a subsequent expansion, non-zero contributions and we relegate this task to the 
code {\tt asy}. The latter yields an output given as a set of $N$-dimensional vectors $\bit{r}_j=\{(r_j)_1,\ldots,(r_j)_N\}$.

In our case of a single small  parameter $t$,  the contribution of a given region $\bit{r}_j$ is obtained from the original integral by adopting the following 
three steps
\begin{align}
\label{MofRrules}
\begin{array}{lll}
& \mbox{(i) rescaling variables as:} & \ x_i \to t^{(\bit{\scriptstyle r}_j)_i} x_i \, , \quad i=1,\ldots,N \, , \\[2mm]
& \mbox{(ii) multiplying integrands by:} & \ t^{\sum_{i=1}^N (\bit{\scriptstyle r}_j)_i} \, , \\[2mm]
& \mbox{(iii) expanding integrands as:} & \ t \to 0 \, .
\end{array}
\end{align}
To see how this works, let us consider an example, say, the three-loop integrals $T_{3,6}$ which is in fact the most complicated case. 
The ``propagators" defining it read
\begin{align}
\mbox{\tt Props}
=
\{
&
-k_1^2, -k_2^2, -k_3^2, -(k_1 + p_1)^2, -(k_1 + k_2 + p_1)^2, -(k_1 + k_2 + k_3 + p_1)^2, 
\nonumber\\
&
-(k_1 + k_2 + k_3 - p_2)^2, -(k_2 + k_3 - p_2)^2, -(k_2 - p_2)^2, -(k_1 + p_1 + p_2)^2
\}
\, ,
\end{align}      
where $k_i$ ($i = 1,2,3$) are the loop momenta. The tenth invariant is in fact a numerator such that our integral $T_{3,6}$ possesses the indices 
$\{a_1, \dots, a_9, a_{10} \} = \{1,\ldots,1,-1\}$. The Feynman integral in question is finite in four dimensions. However, when expanded in the 
limit $t \to 0$ different regions being integrated over the entire $R_+^N$ space inevitably induce divergences. These need to be regularized to 
get finite results. Dimensional regularization comes to the rescue as the most optimal choice and it mends singularities to becomes poles in 
$\varepsilon$. Ultimately, cancellations of the latter becomes then a very powerful check of the correctness of the expansion procedure.

After running {\tt asy} with the help of the {\tt FIESTA}'s command {\tt SDExpandAsy}, we obtain information about all contributing regions.  For the
case at hand there are 35 of them 
\begin{align}
\bit{r}
=
\big\{
&
\{1, 1, 1, 0, 1, 1, 1, 0, 1, 0\}, 
\{0, 0, 0, 0, 0, 0, 0, 0, 0, 0\}, 
\{0, 1, 0, 1, 2, 2, 0, 0, 0, 0\}, 
\nonumber\\
&
\{1, 1, 0, 1, 2, 2, 1, 0, 0, 0\}, 
\{0, 0, 1, 1, 1, 2, 0, 0, 1, 0\}, 
\{0, 1, 1, 0, 1, 1, 0, 0, 1, 0\}, 
\nonumber\\
&
\{0, 1, 0, 0, 1, 1, 0, 0, 0, 0\}, 
\{1, 1, 0, 0, 1, 1, 1, 0, 0, 0\}, 
\{0, 1, 1, 1, 2, 2, 1, 1, 1, 0\}, 
\nonumber\\
&
\{0, 1, 1, 0, 1, 1, 1, 1, 1, 0\}, 
\{0, 1, 1, 1, 2, 2, 0, 0, 1, 0\}, 
\{1, 0, 0, 1, 1, 1, 1, 0, 0, 0\}, 
\nonumber\\
&
\{1, 0, 1, 0, 0, 1, 1, 0, 1, 0\}, 
\{1, 0, 1, 1, 1, 2, 1, 0, 1, 0\}, 
\{0, 0, 0, 0, 1, 1, 1, 0, 0, 0\}, 
\nonumber\\
&
\{0, 0, 0, 1, 1, 1, 0, 0, 0, 0\}, 
\{1, 1, 1, 0, 1, 1, 2, 1, 1, 0\}, 
\{1, 0, 1, 1, 1, 1, 2, 2, 1, 0\}, 
\nonumber\\
&
\{1, 0, 2, 1, 1, 2, 1, 1, 1, 0\}, 
\{1, 0, 1, 1, 1, 2, 2, 1, 1, 0\}, 
\{0, 0, 1, 0, 0, 1, 1, 1, 1, 0\}, 
\nonumber\\
&
\{0, 0, 1, 0, 1, 1, 1, 1, 0, 0\}, 
\{1, 0, 0, 0, 0, 0, 1, 0, 0, 0\}, 
\{0, 0, 1, 1, 1, 2, 1, 1, 1, 0\}, 
\nonumber\\
&
\{0, 0, 1, 0, 0, 1, 0, 0, 1, 0\}, 
\{0, 0, 1, 0, 1, 1, 0, 0, 0, 0\}, 
\{0, 0, 0, 1, 1, 1, 1, 1, 1, 0\}, 
\nonumber\\
&
\{1, 0, 1, 0, 0, 1, 2, 1, 1, 0\},
\{1, 0, 0, 0, 0, 0, 2, 1, 1, 0\}, 
\{1, 0, 1, 1, 1, 1, 1, 1, 0, 0\}, 
\nonumber\\
&
\{1, 1, 1, 1, 2, 2, 1, 0, 1, 0\}, 
\{1, 0, 2, 1, 1, 2, 2, 2, 1, 0\}, 
\{1, 1, 1, 1, 2, 2, 1, 1, 0, 0\}, 
\nonumber\\
&
\{0, 0, 0, 0, 0, 0, 1, 1, 1, 0\}, 
\{1, 0, 0, 1, 1, 1, 2, 1, 1, 0\}
\big\}
\, ,
\end{align}
and they scale, respectively, as the following function of $t$
\begin{align}
\Big\{
&  
t^{-5 \varepsilon},t^0,t^{-6 \varepsilon},t^{-5 \varepsilon},t^{-6 \varepsilon},t^{-2 \varepsilon},t^{-3 \varepsilon},t^{-6\varepsilon},
t^{-4 \varepsilon},t^{-\varepsilon},t^{-5 \varepsilon},
t^{-2 \varepsilon},
\nonumber\\
&
t^{-6 \varepsilon},t^{-5 \varepsilon},t^{-3 \varepsilon}, 
t^{-3 \varepsilon},t^{-4 \varepsilon},t^{-4 \varepsilon},t^{-4 \varepsilon}, t^{-4 \varepsilon},t^{-2 \varepsilon},
t^{-2 \varepsilon},
t^{-3 \varepsilon},
t^{-5 \varepsilon},
\nonumber\\
&
t^{-3 \varepsilon},t^{-3 \varepsilon},t^{-6 \varepsilon},t^{-5 \varepsilon},t^{-6 \varepsilon},
t^{-\varepsilon},t^{-4 \varepsilon},t^{-3 \varepsilon},t^{-4 \varepsilon},t^{-3 \varepsilon},t^{-5 \varepsilon}
\Big\} 
\, .
\end{align}
It becomes immediately obvious that the momentum-space language would make the goal of identification of loop-momentum scalings quite a tedious
task. While it is more or less clear that the $t^0$ behavior is associated with the hard-hard-hard region in the three loop momenta, $t^{-\varepsilon}$ 
is associated with regions where one of the loop momenta is collinear (to $p_i$) and the other two are hard, however, a proper identification of other 
regions in the momentum-space formalism looks quite challenging. Nevertheless, such an analysis will definitely be beneficial since it would result in a 
factorization theorem for the off-shell Sudakov form factor akin to the on-shell case \cite{Korchemsky:1988hd,Collins:1989bt}. 

The analytical mode of the command {\tt SDExpandAsy} provides explicit parametric integrals corresponding to the above 35 contributing regions. To 
systematically evaluate these in the $\varepsilon$-expansion, we rely on the method of the Mellin-Barnes (MB) representation (see, e.g., Ref.\ 
\cite{Smirnov:2012gma}). The latter is based on the simple formula 
\begin{align}
\frac{1}{(A+B)^{\lambda}} = \frac{1}{\Gamma(\lambda)}
\int_{\cal C} \frac{d z}{2 \pi i} 
\frac{B^z}{A^{\lambda+z}} \Gamma(\lambda+z)\Gamma(-z) \; ,
\label{MB} 
\end{align}
which allows one to partition a complicated polynomial in terms of its two `simpler' components $A$ and $B$. In this equation, the contour ${\cal C}$ 
runs from $-i \infty$ to $+i \infty$ in the complex plane and the poles of $\Gamma(\ldots+z)$ are to its left while the ones of $\Gamma(\ldots-z)$ are 
to its right with these left/right poles corresponding to infrared/ultraviolet singularities of the original integral. This formula is usually applied repeatedly 
enough number of times to a given parametric integral in order to transform it into a multiple MB integral. Of course, one attempts to arrive at as simple 
final representation as possible with fewer complex integrations. This procedure was recently automatized with the code {\tt MBcreate.m} \cite{Belitsky:2022gba}.

Once a reasonable MB representation for a given parametric integral is obtained, the next step is to resolve integrand's singularities in $\varepsilon$. 
The goal here is to represent a given complex integral as a linear combination of MB integrals whose $\varepsilon$-expansion can be performed 
{\sl under} the integral sign. There are two public codes {\tt MB.m} and {\tt MBresolve.m} \cite{Czakon:2005rk,Smirnov:2009up}, where this algorithm is 
implemented. They are based on integrations strategies developed in \cite{Tausk:1999vh,Smirnov:1999gc}. We relied on {\tt MBresolve.m}.

The next step is to evaluate emerging MB integrals emerging as the coefficients in the Laurent expansion in $\varepsilon$. Here the command 
{\tt DoAllBarnes} from Kosower's\footnote{All required MB tools can be downloaded from {\tt bitbucket.org/feynmanIntegrals/mb/src/master/}.} 
{\tt barnesroutines.m} automatically applies the first and the second Barnes lemmas (and their corollaries) and thereby 
performs some integrations in terms of the Euler gamma functions. After all these possibilities were exhausted and if some one- and two-fold MB 
integrals are still left, one can turn to numerical evaluations with high accuracy and then apply the {\tt PSLQ} algorithm~\cite{PSLQ:1999} to obtain 
analytic results given a basis of numbers,--- typically values of Riemann zeta function,--- entering the final result is known.
 
The strategy formulated above was successful in our calculation: we obtained asymptotics for all three-loop integrals $T_{3,i}$ and confirmed their 
complete agreement with the results determined via differential equations. Further making use of the numerical mode of {\tt FIESTA}, these findings 
were also verified numerically.

\section{Sudakov form factor: three loops and beyond}
\label{s5}

Having calculated individual contributions in the previous section, we can neatly combine them in linear combinations which arise in the 
three-loop expression (\ref{LogMoffshell3}) and observe massive cancellations of odd powers of the logarithm $\log t$ such that the 
small-$m$ expansions for the $T_{3,i}$ integrals
\begin{align}
\label{TtoDRelation}
Q^6 T_{3,1} & = - \Phi_3 + O(m^2),\nonumber\\
Q^4 T_{3,2} & = \Phi_3 + O(m^2),\nonumber\\
Q^4 [T_{3,3}-T_{3,5}] & = \frac{1}{2}\left[ \Phi_3-\Phi_1\Phi_2 \right] + {O}(m^2),\nonumber\\
Q^4 [T_{3,4}-T_{3,6}] & = -\Phi_1\Phi_2 + {O}(m^2),\nonumber\\
Q^4 T_{3,7} & = \Phi_3 + O(m^2),\nonumber\\
Q^4 T_{3,8} & = \Phi_1\Phi_2 + O(m^2)
\, ,
\end{align}
is determined exclusively by the Davydychev-Usyukina functions $\Phi_\ell \equiv \Phi_\ell (t,t)$. We remind the reader that indeed the small-$m$ expansion 
of $\Phi_\ell$ possesses even powers of $\log t$ one, as can be observed from the explicit expressions for $\Phi_{1,2}$ and $\Phi_3$ previously quoted in 
Eqs.\ (\ref{BoxSmallMassExp1}) and (\ref{BoxSmallMassExp3}), respectively. It is worth pointing out that making use of the symmetry properties of the 
Davydychev-Usyukina functions with respect to their two arguments, the above relations for $T_{3,2}$ and $T_{3,7}$ are similar to the so-called 
``magic identities" of Ref.\ \cite{Drummond:2006rz}. Other relations, however, cannot be obtained in this manner and are therefore unique in this regard.

Substituting (\ref{TtoDRelation}) in $F_2^{(3)}$ and expanding $\log F_2$ in powers of $g$ we found that, up to the three-loop order, $\log F_2$ equals to:
\begin{align}
\label{LogM2octagon}
\log F_2 \left(t,g\right) = -\frac{1}{2} \Gamma_{\rm oct} (g) \log^2 t - D(g)+\mathcal{O}(m^2),
\end{align}
with the functions $\Gamma_{\rm oct} (g)$ and $D(g)$ of the coupling quoted earlier in Eqs.\ (\ref{ExactGammaOct}) and (\ref{ExactDet}), respectively. 
This is exactly the logarithm of the null octagon $\mathbb{O}_0(z,\bar{z})$ \cite{Coronado:2018cxj,Belitsky:2019fan} multiplied by the factor of $2$ and 
expanded up to $O (g^6)$,
\begin{align}
\log\mathbb{O}_0(z,\bar{z})
=
-\frac{1}{4} \Gamma_{\rm oct}(g) \log^2\left(\frac{\bar z}{z}\right)
- g^2 \log(z\bar z)-\frac{1}{2} D(g)
\, ,
\end{align}
with $z = 1/\bar{z} = \sqrt{t}$. It appears natural to us to conjecture that this relation holds at any order of perturbation theory as well, i.e., 
\begin{align}
\label{LogM2conjcture}
\log F_2 = 2\log \mathbb{O}_0 + O(m^2),
\end{align}
and thus conjecturing that the off-shell Sudakov form factor is given by the null octagon function $\mathbb{O}_0$ to all orders of the perturbation 
theory. This is the main result of our work.

Let us make some comments regarding our claim. We see that all integrals (individual integrals or their linear combinations) in (\ref{LogMoffshell3}) 
can be represented as linear combinations of products of Davydychev-Usyukina function $\Phi_{\ell}(t,t)$ at least up to $O(m^2)$ terms. This observation 
is reminiscent of the results of \cite{Caron-Huot:2021usw} where all scalar integrals contributing to four point amplitude in off-shell kinematical regime 
where expressed in terms of linear combinations of products of $\Phi_{\ell}(x,y)$ functions up to four loops. The reason why such relations between integrals 
exist can be traced back to the integrability of underlying problem. The four point amplitude in off-shell kinematical regime is expected to be given by 
$\mathbb{O}_0(z,\bar{z})$, which can be written in closed form, for instance as a perturbative series \cite{Coronado:2018cxj,Kostov:2019stn,Kostov:2019auq},
\begin{align}
\label{OctDef1}
\mathbb{O}_0=\det (1-\mathbb{K}_0)\, , \qquad
(\mathbb{K}_0)_{nm}=\sum_{\ell=n+m-1}^{\infty}(- g^2)^\ell C_{nm}^{(\ell)} \Phi_{\ell}(z,\bar z),
\end{align}
where $C_{nm}^{(\ell)}$ are explicitly known coefficients
\begin{equation}
C_{nm}^{(\ell)}=\frac{-(2m-1)[2\ell]![\ell-1]! \ell !}{[\ell-(n+m-1)]![\ell+(n+m-1)]![\ell-|n-m|]![\ell+|n-m|]!}
\, .
\end{equation}
The relations (\ref{TtoDRelation}) together with (\ref{12loopIntegralsPlanar}) and (\ref{2loopIntegralNPlanar}) can be considered as subtle hints that the 
Sudakov problem (for the two-particle form factor, in the current case, or generally even multi-leg ones) in $\mathcal{N}=4$ sYM on the Coulomb branch 
in the off-shell kinematical regime can potentially be solved using integrability. Another hint pointing to this conclusion is that Eqs.\ (\ref{TtoDRelation}),
(\ref{12loopIntegralsPlanar}) and (\ref{2loopIntegralNPlanar}) express nonplanar integrals in terms of planar one. The latter in turn possess well defined 
dual conformal symmetry properties,  which being a part of a larger symmetry group,--- the so-called Yangian symmetry \cite{Drummond:2009fd} ,---
is intrinsic to integrable systems, see, e.g., \cite{Loebbert:2016cdm}.

\section{Conclusions}

As a conclusion, let us make several comments regarding IR properties of the off-shell Sudakov form factor and amplitudes which we observed as a result
of our analysis. The most obvious one is that in the off-shell regime the Sudakov logarithms indeed exponentiate but it is $\Gamma_{\rm oct}$ rather
than $\Gamma_{\rm cusp}$ that governs the rate of its decay. As was already anticipated in \cite{Caron-Huot:2021usw}, the $\Gamma_{\rm cusp}$ is not 
the archetypal IR anomalous dimension and IR behavior of amplitudes and form factors is far more involved than previously expected. Note also that 
in the off-shell case there are no $\log^{2n+1}$ terms and hence there is no analogue of the collinear anomalous dimension, at least, up to the three loop 
accuracy. In terms of evolution equation, the result (\ref{LogM2octagon}) can be rewritten as:
\begin{align}
\label{EvolEq1}
\left( \frac{\partial}{\partial \log m^2} \right)^2 \log F_2=-\Gamma_{\rm oct} (g)
\, .
\end{align}
This evolution equation is obviously different from what was conjectured earlier in the literature, e.g., \cite{Drummond:2007aua}, which involved
$\Gamma_{\rm cusp}$ instead. 

Let us point out that factorization properties for four- and five-leg amplitudes for W-boson scattering still hold in this off-shell kinematical regime
such that IR-sensitive parts are still driven by the product of Sudakov form factors identical to (\ref{AmplFactor}), but now with $F_2$ given by Eq.\
(\ref{LogM2octagon}). This demonstrates self-consistency of these considerations in such kinematical regime. A natural extension of these findings
would be to study the structure of the Sudakov form factor in a similar kinematical regime in QCD and other four-dimensional gauge theories. 
Regarding QCD, our computation can be considered as a determination of its ``most transcendental" part in the planar limit.

In spite of the fact that the form of the evolution equations differ depending on the value of the external particles' off-shellness, (\ref{EvolEq}) vs.\ 
(\ref{EvolEq1}), there is a chance that they are in fact next of kin. The two anomalous dimensions can be found as solutions to the so-called 
flux-tube equations, given in Ref.\ \cite{Beisert:2006ez} for $\Gamma_{\rm cusp}$ and \cite{Belitsky:2019fan} for $\Gamma_{\rm oct}$. The two can 
in fact be combined into a more general equation by introducing a deformation parameter \cite{Basso:2020xts}. Thus, it appears that the latter encodes 
a very subtle, anomalous effect of the non-commutativity of $p_i^2\rightarrow 0$ and $\varepsilon \rightarrow 0$ limits. At the moment we have no 
adequate understanding of this fact. 

\begin{acknowledgments}
We are grateful to A.F.~Pikelner for collaboration. L.B.\ is grateful to A.I.\ Onishchenko for useful discussions and to A.V.\ Bednyakov, 
N.B.\ Muzhichkov and E.S.\ Sozinov for collaboration at early stages of the project. The work of A.B.\ was supported by the U.S.\ National Science
Foundation under the grant No.\ PHY-2207138. The work of L.B.\ was supported by the Foundation for the Advancement of Theoretical Physics and 
Mathematics ``BASIS".  
The work of V.S. was supported by the Russian
Science Foundation under the agreement no. 21-71-30003 (expanding Feynman integrals with expansion by regions) and by the Ministry of Education and Science of the
Russian Federation as part of the program of the Moscow Center for Fundamental and Applied
Mathematics under Agreement No. 075-15-2019-1621 (evaluating contributions of regions with MB representations).
\end{acknowledgments}

\newpage
\appendix

\section{Small off-shellness expansion of $T_{3,i}$}
\label{a2}

In this Appendix, we present a list of the small-$m$ expansions for all $T_{3,i}$ integrals given in terms of HPLs in the previous Appendix.
They read
\begin{align}
Q^6T_{3,1}
&
=- \frac{1}{36}\log^6 t-\frac{5\zeta_2}{6}\log^4 t-\frac{35\zeta_4}{2}\log^2 t-\frac{155 \zeta_6}{4}
\, , \\
Q^4T_{3,2}
&
=\frac{1}{36}\log^6 t+\frac{5\zeta_2}{6}\log^4 t+\frac{35\zeta_4}{2}\log^2 t+\frac{155 \zeta_6}{4}
\, , \\
Q^4T_{3,3}
&
=-\frac{1}{36}\log^6 t+\frac{\zeta_2}{3}\log^4 t+\frac{2\zeta_3}{3}\log^3 t+\frac{27 \zeta_4}{2}\log^2 t
+
\left(4 \zeta_2 \zeta_3 - 20 \zeta_5\right)\log t + 32\zeta_6 - 4\zeta_3^2
\, , \\
Q^4T_{3,4}
&
=-\frac{1}{36}\log^6 t-\frac{5\zeta_2}{6} \log^4 t-\frac{4\zeta_3}{3}\log^3 t+\frac{\zeta_4}{2}\log^2 t
-20\zeta_5 \log t-\frac{27\zeta_6}{4}-16\zeta_3^2
\, , \\
Q^4T_{3,5}
&
=\frac{1}{12}\log^6 t+\frac{5 \zeta_2}{3}\log^4 t+\frac{2\zeta_3}{3}\log^3 t+\frac{35\zeta_4}{2}\log^2 t
+ \left(4 \zeta_2 \zeta_3-20 \zeta_5\right)\log t+31\zeta_6-4\zeta_3^2
\, , \\
Q^4T_{3,6}
&
=\frac{2 }{9}\log^6 t+\frac{8 \zeta_2}{3} \log^4 t-\frac{4 \zeta_3}{3} \log^3 t
+26\zeta_4\log^2 t -20\zeta_5\log t+30 \zeta_6-16 \zeta_3^2
\, , \\
Q^4T_{3,7}
&
=\frac{1}{36}\log^6 t+\frac{5\zeta_2}{6} \log^4 t+\frac{35\zeta_4}{2} \log^2 t+\frac{155 \zeta_6}{4}
\, , \\
Q^4T_{3,8}
&
=\frac{1}{4}\log^6 t+\frac{7\zeta_2}{2}\log^4 t+\frac{51\zeta_4}{2} \log^2 t+\frac{147 \zeta_6}{4}
\, ,
\end{align}
and valid up to $O(m^2)$.

\bibliographystyle{JHEP} 

\bibliography{literature}

\end{document}